\documentclass{ws-ijmpe}
\begin{document}

\markboth{Sayantan Sharma}{Recent theoretical developments on QCD matter at
finite temperature and density}

\catchline{}{}{}{}{}

\title{Recent theoretical developments on QCD matter at
finite temperature and density}

\author{Sayantan Sharma}

\address{The Institute of Mathematical Sciences, HBNI\\
Taramani, Chennai 600113 India\\
sayantans@imsc.res.in}

\maketitle


\begin{abstract}
QCD matter at finite temperature and density is a subject that has witnessed very 
impressive theoretical developments in the recent years. In this review I will 
discuss some new insights on the microscopic degrees of freedom of the QCD 
medium near the chiral crossover transition from lattice QCD. Latest high 
precision lattice data on the fluctuations and correlations between conserved 
charges like the baryon number, strangeness can help us to understand and 
distinguish between different models of interacting hadrons. Furthermore, 
the latest constraints on the location of the critical end-point and the 
curvature of the critical line will be discussed. In the later part of this 
review I will discuss about the insights on the thermal nature of the medium 
created in heavy ion collision experiments that have come from the theoretical 
analysis of the particle yields, and to what extent the lattice data on 
correlations and fluctuations of conserved charges can give us any information 
about the fireball at freezeout. 
\end{abstract}

\keywords{QCD at finite temperature and density, phase transitions, critical end-point}

\ccode{PACS numbers: 1.15.Ha, 12.38.Gc, 12.38.Mh}


\section{Introduction}
The phase diagram of strongly interacting matter described by 
Quantum Chromodynamics (QCD) is intriguing and its complete 
understanding lies at the core of explaining many fundamental 
phenomena like the origin of mass of the visible universe, the 
dynamics of the neutron star mergers, to name a few. It has 
inspired several large-scale experimental programs, the 
Relativistic Heavy-ion collider (RHIC) at Brookhaven National 
Laboratory, the ALICE at CERN and also the 
upcoming ones at FAIR, GSI and at NICA, Dubna. These fantastic 
experiments are designed to explore different regimes of the QCD 
phase diagram in the temperature, $T$ and baryon density $\mu_B$ 
plane. In particular the ongoing BES-II runs at the RHIC, among 
other goals, are searching for an important landmark in the phase 
diagram, the critical end-point of the chiral crossover transition.   
Theoretical predictions starting from the fundamental quantum field 
theory i.e. QCD  is immensely challenging as it is driven by 
non-perturbative interactions. Lattice gauge theory is the only 
available non-perturbative tool that has given us the initial 
breakthrough in predicting the phase diagram at $\mu_B\to 0$ \cite{Schmidt:2017bjt}. 
In this work, I will review how improved lattice techniques 
at finite $\mu_B$, have provided new results on the Equation of 
State (EoS) for $\mu_B/T \lesssim 2.5 $, and the different fluctuations 
and correlations of conserved quantum charges. These results are 
not only providing fundamental insights on the non-perturturbative 
aspects of QCD but are also important inputs towards understanding and 
modeling of the heavy-ion experiments. 

\section{QCD Equation of State at finite density}

Conventional Monte Carlo methods fail to be applied at finite 
baryon density due to the infamous \emph{sign problem}. Serious 
efforts are ongoing within the lattice community to solve this 
problem at least for simple quantum field theories, for a recent 
review see Ref. \refcite{deForcrand:2010ys}. In absence of an immediate 
solution, two methods proposed to circumvent this problem have 
provided most reliable results in finite density QCD in the 
continuum and thermodynamic limit. The first among them is to 
calculate the partition function at imaginary $ \mu_B$ where the 
sign problem is absent and then analytically continuing to the 
real $ \mu_B$ plane \cite{DElia:2002tig}. The other method 
which would be relevant for most of the later discussions in 
this review on thermodynamics at finite density is the Taylor 
series method \cite{Allton:2002zi,Gavai:2003mf}. The idea is 
to calculate the thermodynamic quantities such as pressure as 
a Taylor series in $\mu_B/T$ about $\mu_B=0$.

\begin{equation}
\frac{P(\mu_B,T)}{T^4}=\frac{P(0,T)}{T^4}+\frac{1}{2}\left(\frac{\mu_B}{ T}\right)^2
\frac{\chi_2^B(0,T)}{ T^2}+\frac{1}{4}\left(\frac{\mu_B}{ T}\right)^4\frac{\chi_4^B(0)}{3!}+...
\end{equation}

\begin{figure}[th]
\includegraphics[width=0.3\textwidth]{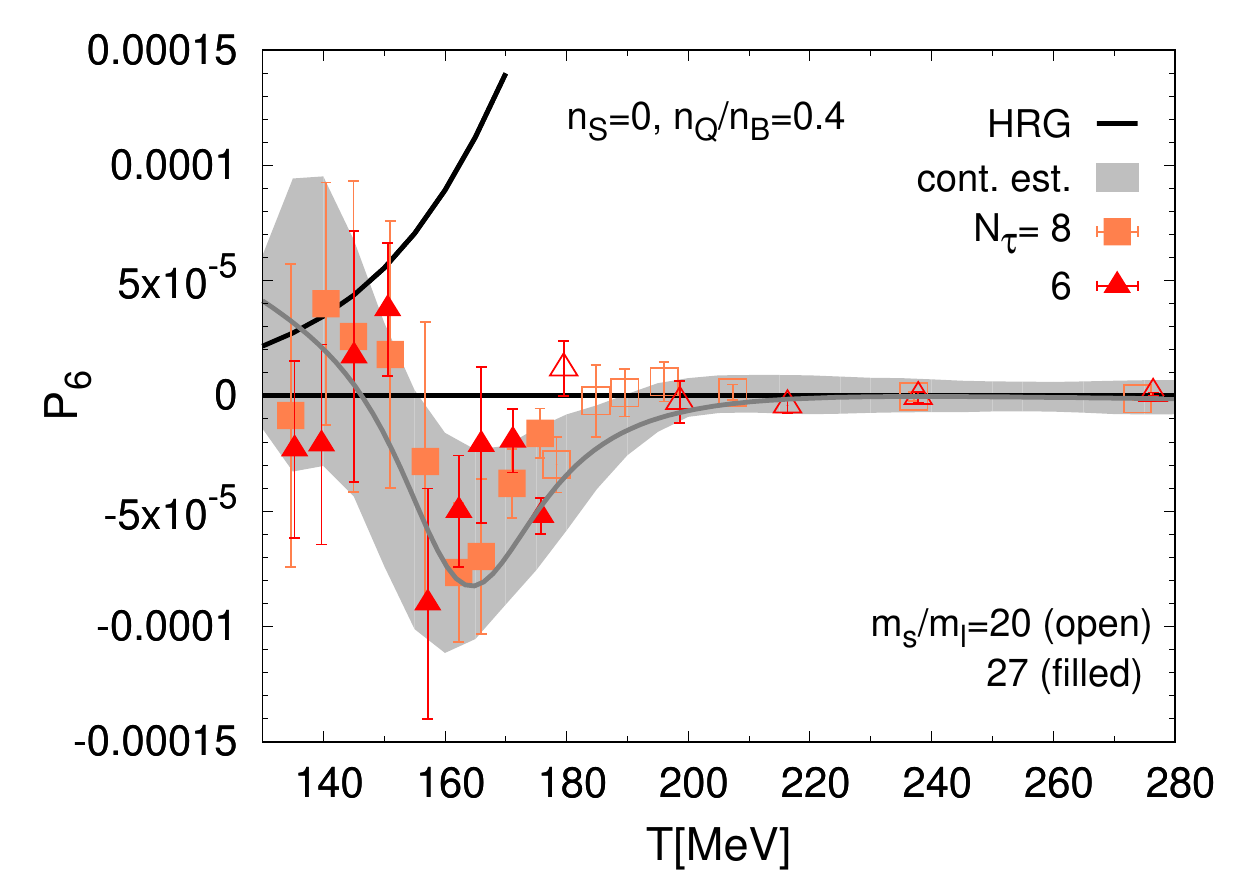}
\includegraphics[width=0.3\textwidth]{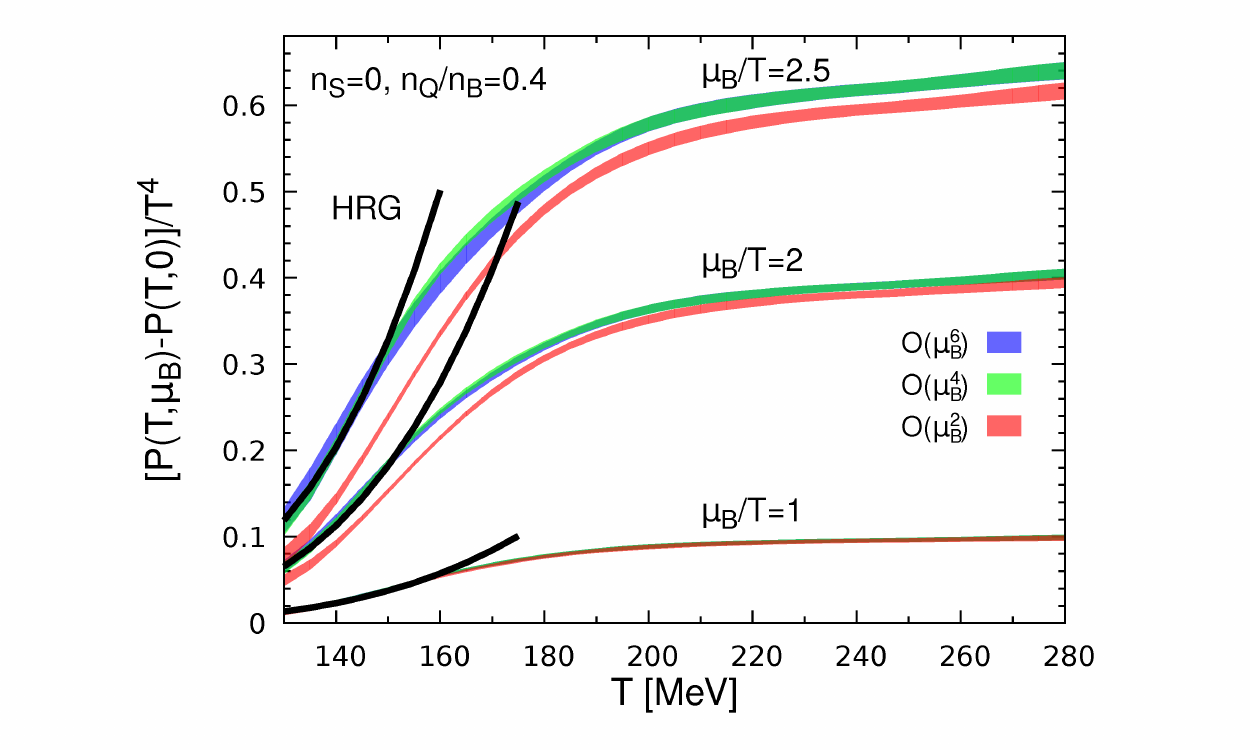}
\includegraphics[width=0.3\textwidth]{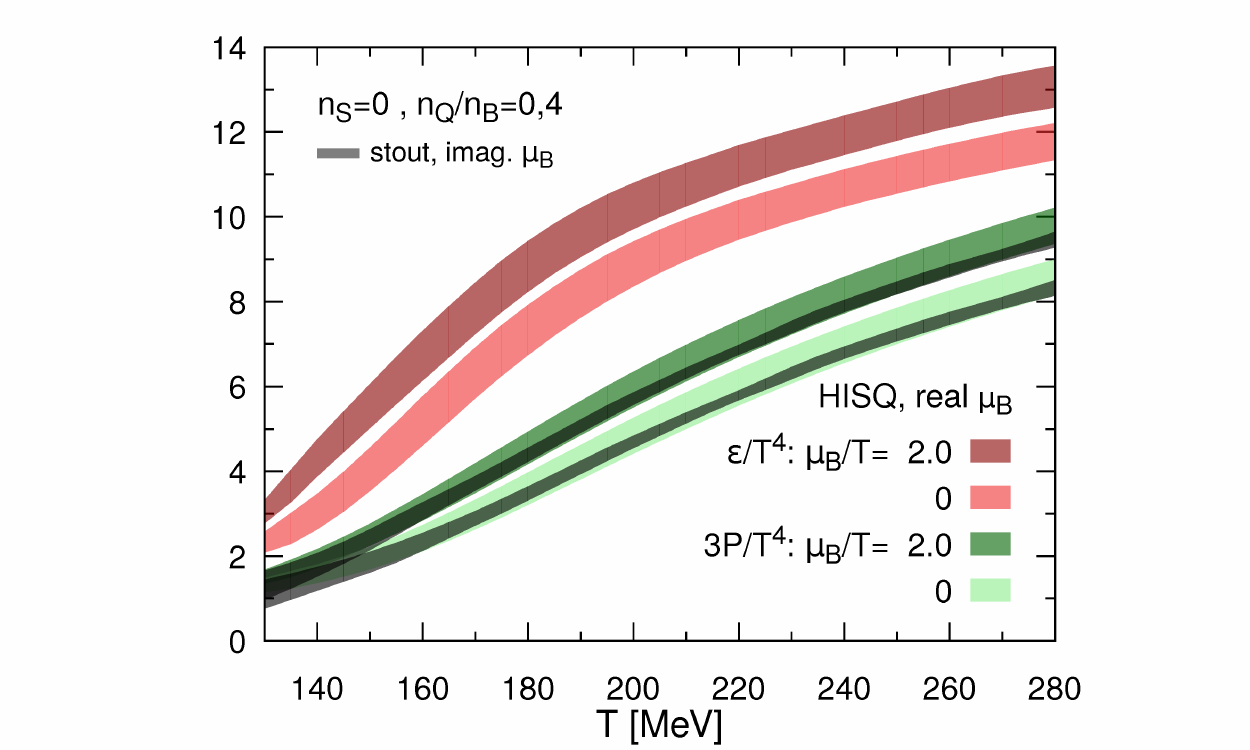}
\caption{The left most panel shows the Taylor coefficient $P_6$ as
a function of temperature for 2+1-f QCD with HISQ. The pressure 
and the energy density for different $\mu_B/T$ for the strangeness 
neutral case are shown in the middle and right panels respectively. }
\label{fig:1}
\end{figure}

All the Taylor coefficients $P_{2n}=\frac{\partial^{2n} P}{(2n)!\partial \mu_B^{2n}}$ 
are measured at zero density without encountering the sign problem. In experiments 
the net strangeness is zero i.e., $n_S=0$ and typically the ratio of net baryon number 
to net charge $n_B/n_Q=0.4$ \cite{Bazavov:2012vg}. These are known as the strangeness 
neutrality constraints. 
The lattice results discussed in this section and in the last section have been obtained 
with this constraint implemented. The continuum estimates of $P_{2,4,6,8}$ have been 
measured on the lattice \cite{Bazavov:2017dus,Borsanyi:2018grb,Bazavov:2020bjn}, 
the temperature dependence of $P_6$ \cite{Bazavov:2017dus} is shown in the left 
panel of Figure \ref{fig:1}. It exhibits distinct non-perturbative features like 
the negative values just above $T_c$ which cannot be explained within the Hard 
Thermal loop perturbation theory.   

As evident from the central panel of Figure \ref{fig:1} that the pressure for 
$\mu_B/T \lesssim 2.5$ is already very well determined using expansion up to 
$\mathcal O(\mu_B^6)$ \cite{Bazavov:2017dus}. To determine the pressure 
at $\mu_B/T\sim 3$ or beyond one needs to calculate $\chi^B_8$ with better 
precision. The right panel shows the continuum estimates of energy density 
and pressure calculated at $\mu_B/T=0,2$ using two different lattice fermion 
dicretizations namely the Highly improved staggered quarks (HISQ) 
\cite{Bazavov:2017dus}, shown as colored bands and the stout improved 
staggered quarks \cite{Gunther:2016vcp}, shown as gray bands taken 
from Ref. \refcite{Bazavov:2017dus}. 
Furthermore HISQ results were obtained using Taylor series method whereas 
the stout results are calculated in the imaginary $\mu_B$ formalism. 
The consistency between different lattice results irrespective of the 
systematics show that indeed the EoS of QCD for $\mu_B/T \lesssim 2.5$ 
has been reliably measured.

\section{Degrees of freedom near $T_c$}
An important component of any phase diagram is understanding how 
the degrees of freedom change as one undergoes a phase transition.
For a strongly interacting system a quasi-particle description may 
not be even possible a priori. This is particularly evident from the 
data for the screening mass in the meson sector which do not 
approach its perturbative values even at $\sim 8~T_c$ \cite{Bazavov:2019www}. 
In this section I will particularly focus of the status of our understanding 
of the warm QCD medium at $T<T_c,~\mu_B=0$. Inspired by the Hagedorn model, 
remarkable proposals have been made towards understanding the properties 
of the chiral symmetry broken phase of QCD in terms of Hadron Resonance 
gas (HRG) model. The main rationale behind constructing such a model is 
when the interactions between hadrons proceed via resonant channels, an 
interacting system of hadrons can be simply mimicked by a non-interacting 
gas of hadrons and all possible resonances \cite{Dashen:1969ep,Venugopalan:1992hy}.
With very high precision lattice QCD data available, this ansatz can be and 
is put into rigorous scrutiny. Already it has been observed that the a naive 
HRG model constructed out of all experimentally known hadron states including 
resonances from the Particle Data Group cannot be a good description of the 
thermodynamic properties of the strange \cite{Bazavov:2014xya} and charm 
baryons \cite{Bazavov:2014yba}. Many new lattice studies in the last few years 
have demonstrated the failure of a naive HRG model and how to improve upon them. 
The main results are summarized below,
\vspace{-0.3cm}      
\begin{romanlist}[(ii)]
\item Many  (multi) strange and charm hadrons and resonances have not 
been yet detected in the experiments but predicted from lattice studies 
and quark models \cite{Bazavov:2014xya,Bazavov:2014yba}, which contributes 
to thermodynamics near $T_c$.
\item  The baryon interactions may not be always resonant and can be repulsive.
This results in the breakdown of the naive HRG description of the higher 
order cumulants of baryon number like $\chi_6^B$ \cite{Huovinen:2017ogf}.
\item The thermal broadening of hadron masses have to be taken into account. 
For e.g., negative parity baryon masses receive significant thermal modifications 
compared to their parity partners and they become degenerate at $T_c$ 
\cite{Aarts:2017rrl,Aarts:2018glk}.
\end{romanlist}

\subsection{Constraining the HRG from latest lattice data}

\begin{figure}[th]
\centerline{\includegraphics[width=0.3\textwidth]{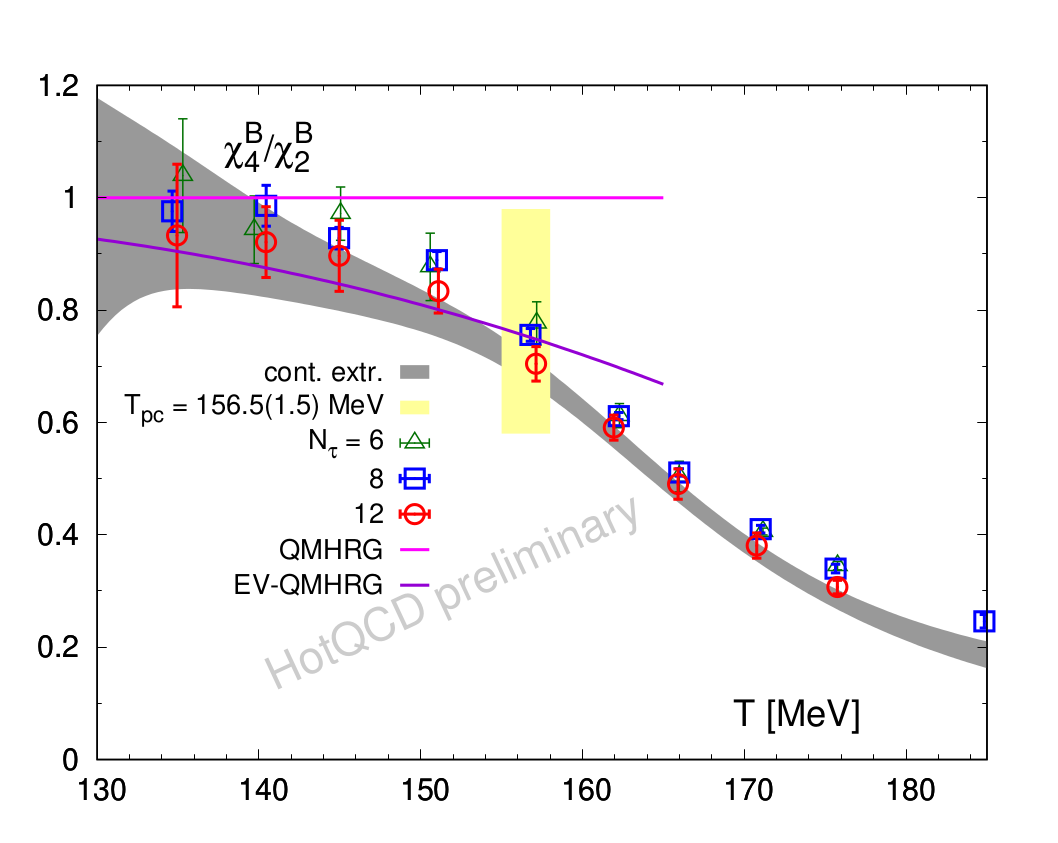}
\includegraphics[width=0.3\textwidth]{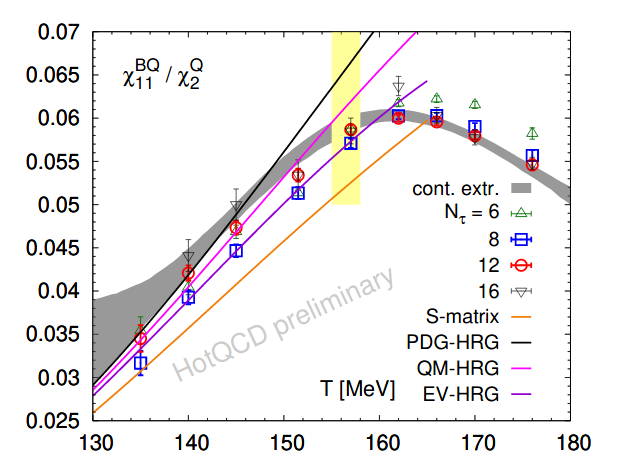}
\includegraphics[width=0.3\textwidth]{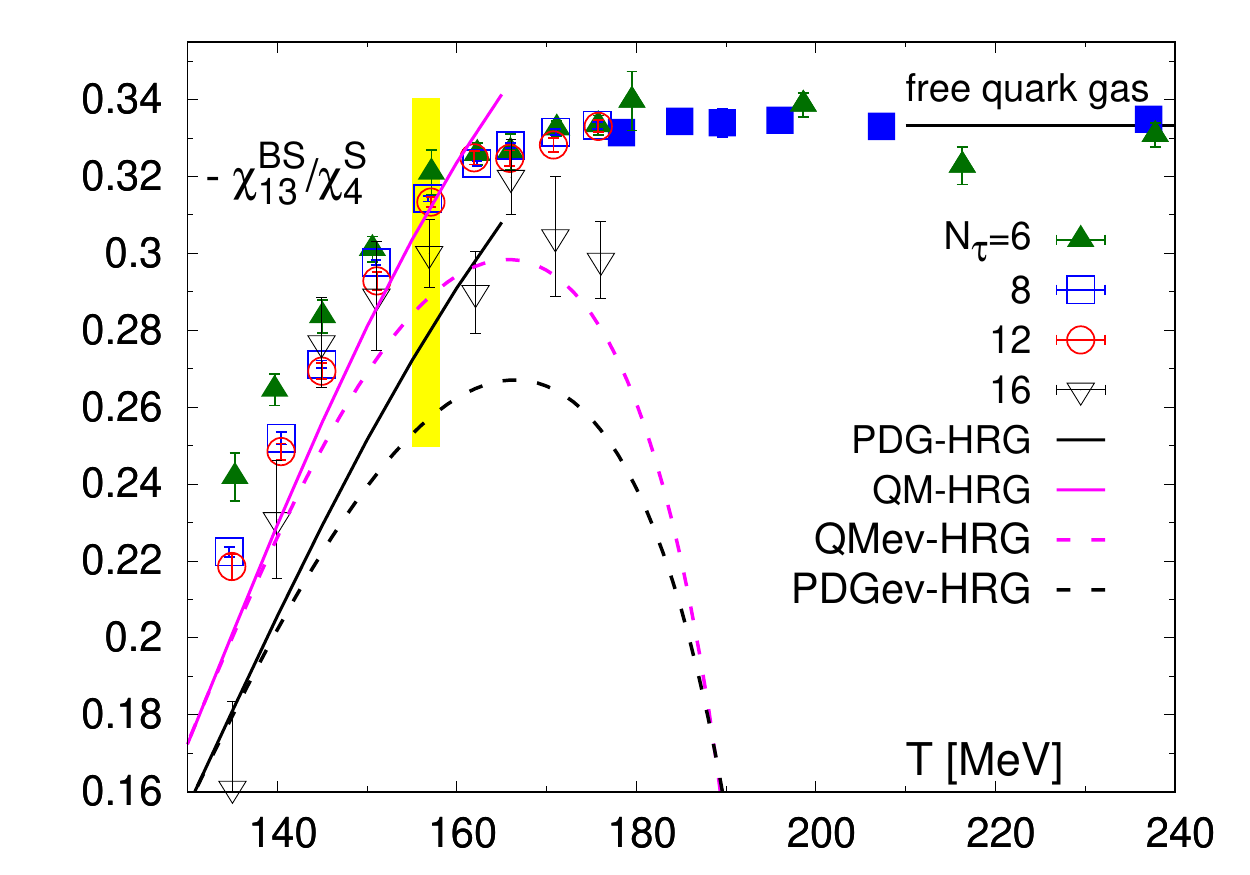}
}
\caption{The different ratios of cumulants of conserved charges in 
QCD calculated using lattice techniques compared to the predictions 
from different Hadron Resonance gas models. }
\label{fig:2}
\end{figure}

In recent years, the availability of very high precision lattice data
on various cumulants of conserved charges have allowed for better 
constraining of HRG models. For example the continuum results for the 
ratio $\chi_4^B/\chi_2^B$, shown in the left panel of Figure \ref{fig:2} 
already favors including repulsive interactions within the baryon sector 
of HRG at mean field level via excluded volume 
\cite{Bazavov:2020bjn,Rischke:1991ke,Goswami:2020yez}.  

However central panel of Figure \ref{fig:2} demonstrates such a mean-field 
treatment of the baryon interactions may not be adequate in explaining the 
temperature dependence of the $\chi_{11}^{BQ}/\chi_2^Q$ \cite{Goswami:2020yez}. 
In the strangeness sector the importance identifying the resonances and their 
interactions is emphasized through the comparison of ratios like 
$\chi_{13}^{BS}/\chi_4^S$, shown in right panel of Figure \ref{fig:2} 
\cite{Goswami:2020yez}. 
Different smart combinations of correlations of strangeness with other 
quantum numbers have been proposed \cite{Goswami:2020yez} for a better 
understanding of the thermodynamics of multi-strange baryons close 
to $T_c$. Constraining the HRG description driven by the lattice data 
on thermal QCD, will ultimately help us to use such models in understanding 
the dynamical properties of QCD, where lattice techniques are yet to be 
developed. 

\section{The pseudo-critical line and the critical end-point }

\begin{figure}[bh]
\vspace{-0.5cm}
\centerline{
\includegraphics[width=0.55\textwidth]{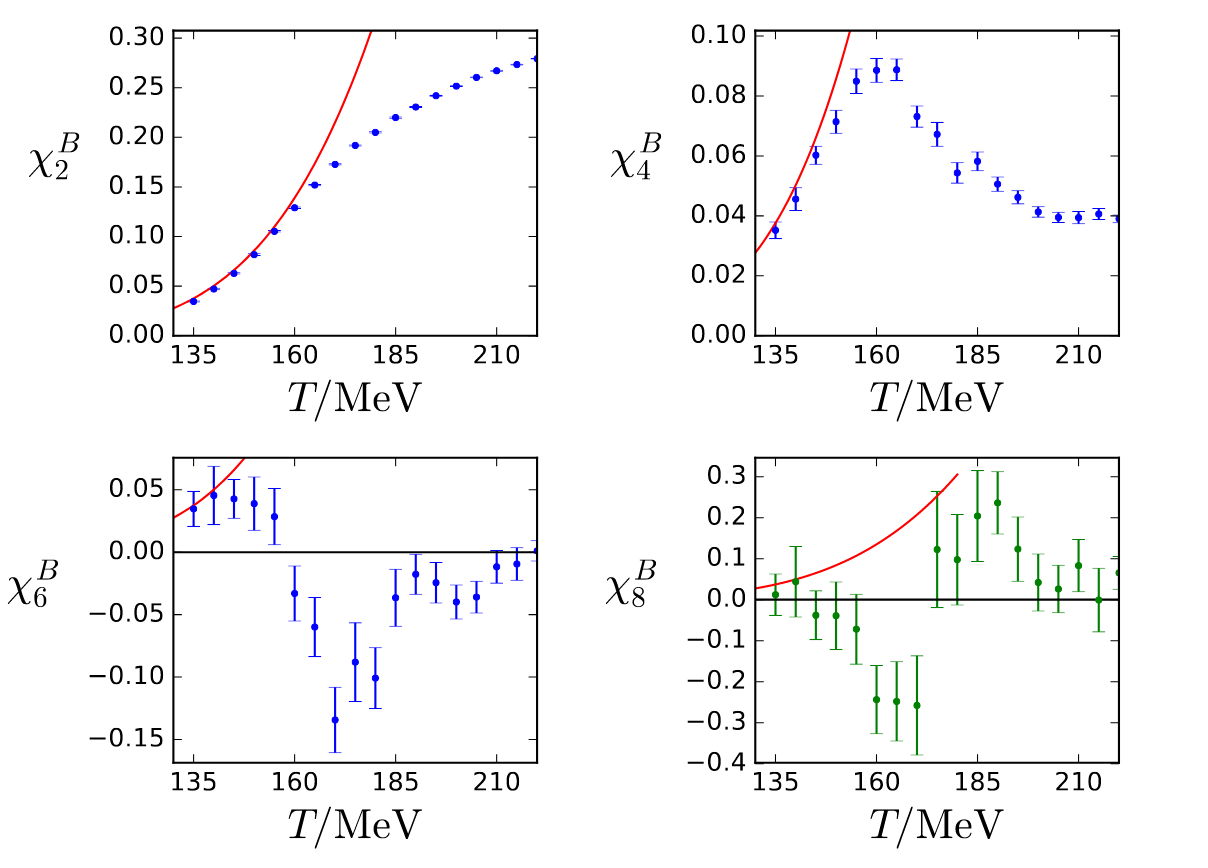}
\includegraphics[width=0.25\textwidth]{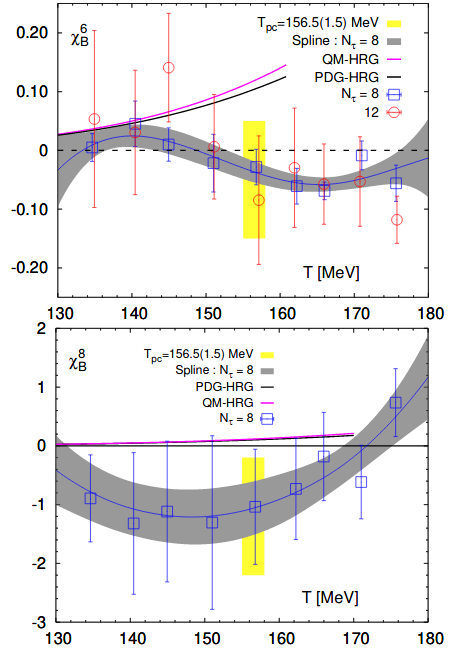}
}
\caption{The continuum estimates of the baryon number fluctuations 
$\chi_n^B$ obtained from recent high-statistics lattice studies. 
The left panel shows the continuum estimates for $n=2$-$8$ obtained 
from $2+1$ flavor QCD with stout staggered fermion discretization 
and $N_\tau=12$. The right panel shows the latest continuum estimates 
for $\chi_{6,8}^B$ for $2+1$ QCD with HISQ discretization.}
\label{fig:3}
\end{figure}

Moving closer to the chiral crossover transition, we are now progressing 
towards charting out the pseudo-critical line in the $T$-$\mu_B$ plane.  
Mathematically this is represented as 
$\frac{T_c(\mu_B)}{T_c(0)}=1-\kappa_2^B \frac{\mu_B^2}{T_c(0)^2}-\kappa_4^B \frac{\mu_B^4}{T_c(0)^4}-... $,
where $\kappa_n$ are the n-th order curvature coefficients. Different 
lattice groups have now converged on the final result for $\kappa_{2}^B\sim 0.01$ 
using different techniques with an error $<20\%$,  
\cite{Bonati:2018nut,Bazavov:2018mes,Borsanyi:2020fev} and 
$\kappa_{4}^B\sim 0$ \cite{Bazavov:2018mes,Borsanyi:2020fev} albeit with
a large error due to systematic uncertainties in performing a continuum 
extrapolation of this observable.

\begin{figure}[h]
\centerline{
\includegraphics[width=0.45\textwidth]{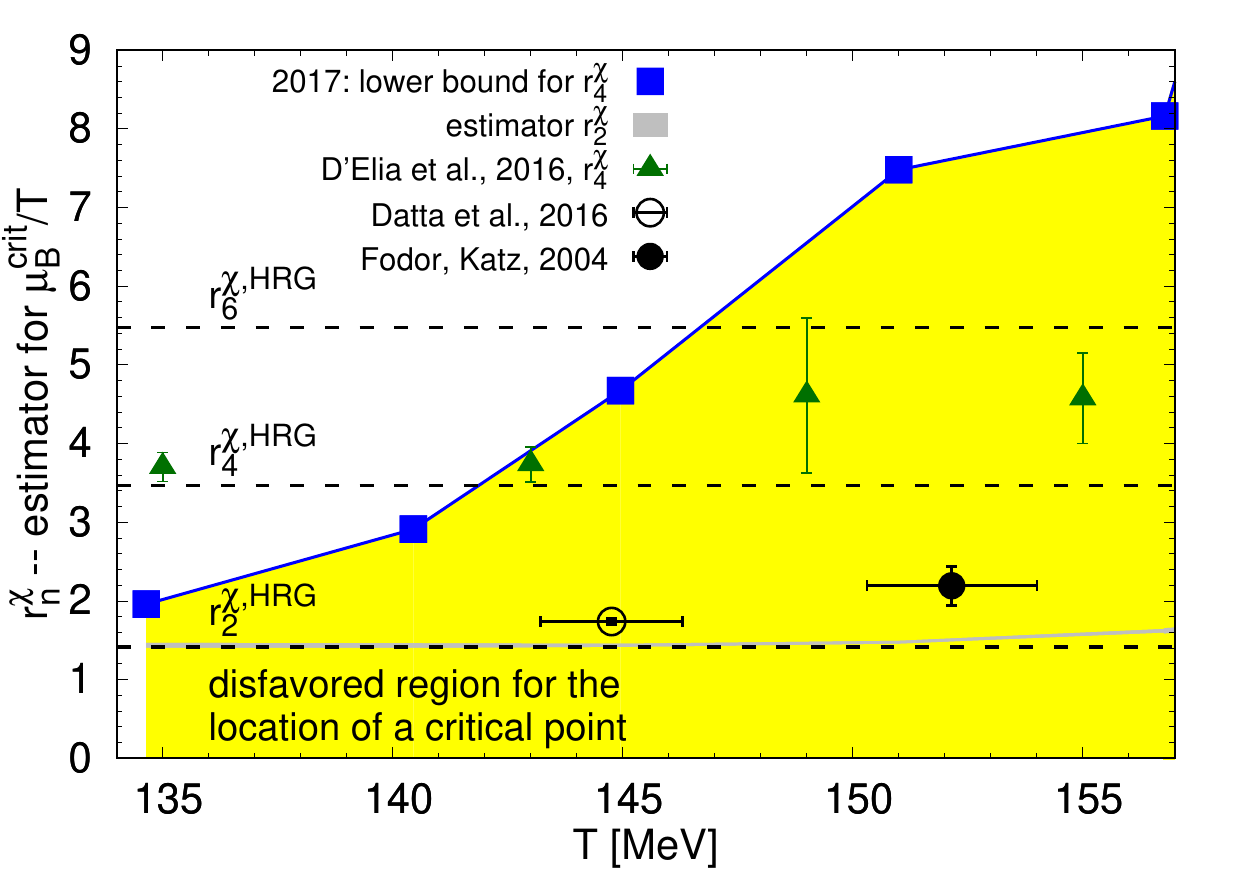}
\includegraphics[width=0.45\textwidth]{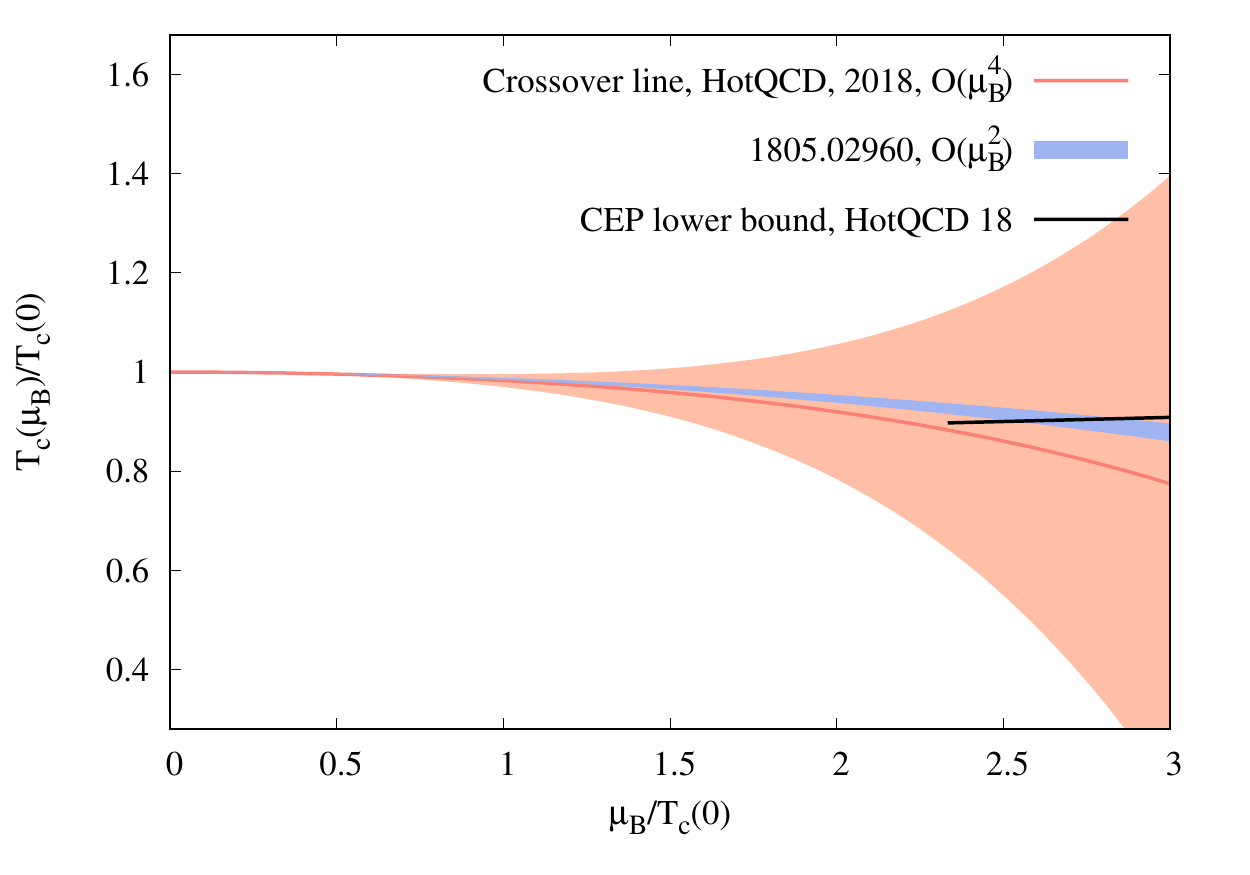}
}
\caption{A summary of the radius of convergence estimates 
for pressure as a function of temperature on finite lattices, 
is shown in the left panel. 
The right panel shows the current status of the QCD phase diagram and 
the CEP estimates shown as a black line vis-a-vis the pseudo-critical line. }
\label{fig:4}
\end{figure}

From model studies it is conjectured that at large $\mu_B$, the chiral 
phase transition should be first order hence this crossover line should 
end at a critical end-point (CEP). The location of the CEP can be 
estimated from the radius of convergence of thermodynamic quantity 
such as pressure as a function of $\mu_B/T$ and is defined as 
$\lim_{n \to \infty}r_{2n} \equiv 
\sqrt{2n(2n-1)\vert\frac{\chi_{2n}^B}{\chi_{2n+2}^B} \vert}$.
Practically it is not known how large this $n$ should be chosen on 
the lattice. To have a rough estimate it is instructive to look at 
the temperature dependence of the $\chi_{2n}^B$ measured at $\mu_B=0$.
Data are currently available for up to $2n=8$ for physical quark masses, 
which are obtained from the imaginary $\mu_B$ technique \cite{Borsanyi:2018grb}, 
shown in the left panel of Figure \ref{fig:3} compared to those obtained using 
Taylor expansion \cite{Bazavov:2020bjn}, shown in the right panel 
of the same Figure. In order to have a CEP in the real $\mu_B$ plane, the 
successive $\chi_{2n}^B$ have to be all positive which imposes a constraint on 
$T_{CEP} \lesssim 135$ MeV.

Radius of convergence estimates, $r_{2n}$ are summarized 
in the left panel of Figure \ref{fig:4}. These estimates are for 
finite lattice spacings and volumes with data taken from Refs. 
\refcite{Fodor:2004nz,DElia:2016jqh,Datta:2016ukp,Bazavov:2017dus}.
Though a general consensus will emerge only when thermodynamic and 
continuum extrapolations are performed, the preliminary high statistics 
results using the Taylor expansions and measured with HISQ discretization 
already constrain the location of the CEP at $\mu_B/T>2.5$ for 
$135 \leq T< 140$ MeV \cite{Bazavov:2017dus}.  Latest understanding of the 
phase diagram near CEP is summarized in the right panel of Figure 
\ref{fig:4}. It is clearly evident that the NLO curvature coefficient 
$\kappa_4^B$ will be crucial to constrain the location of CEP. If 
$\kappa_4^B=\kappa_6^B\sim0$ then the pseudo-critical line will not 
bend significantly and one can estimate the CEP already with $\chi_8^B$. 
On the other hand $\kappa_4^B>0$ will increase the curvature of the 
pseudo-critical line as a function of $\mu_B$, thus one would require 
estimates beyond $\chi^B_8$ to reliably locate the CEP.

\section{Can lattice data be meaningfully used for understanding 
chemical freezeout?}

The fireball created during heavy-ion collisions expands 
quite rapidly and hadronizes within $10$ fm/c. A wealth of 
data from the RHIC and ALICE suggests that the system achieves early 
hydrodynamization \cite{Gale:2013da}. Hydrodynamic simulations have 
very successfully explained several properties of the fireball 
\cite{Gale:2013da} and most use the thermodynamic EoS from the 
lattice. Thus dynamical modelling of the fireball suggests
existence of a local thermodynamic equilibrium. The process of 
hadronization and chemical freezeout is not yet very well 
understood i.e. if the conditions at the chemical freezeout 
can be described by a uniform $T_f$ and $\mu_B$. 

The experimental yields of baryons and mesons, taking into account the 
feed-down from higher excited states, have been fitted to their 
corresponding values within the HRG which is characterized by three 
parameters $\mu_B~,T_f$ and the volume of the fireball. Such fits 
have been largely successful, explaining yields of several species 
of baryons and mesons \cite{Andronic:2005yp,Cleymans:2005xv}. However 
recent attempts of such fits on the latest ALICE data from central Pb-Pb 
collisions at $\sqrt s=2.76$ TeV have thrown new surprises \cite{Andronic:2017pug}. 
Though statistical fits can explain most of the yields and predicts a 
$T_f\sim 156$ MeV and $\mu_B\sim 0$ \cite{Andronic:2017pug}, a glaring 
exception is the fit to the experimental proton yield which deviates 
significantly by $2.6~\sigma$ \cite{Andronic:2018qqt}. Refining the HRG 
model by including the additional states predicted from quark models and 
lattice, could not reduce this discrepancy \cite{Andronic:2018qqt}. A 
recent extensive study have demonstrated \cite{Andronic:2020iyg} that 
$N^{*}, \Delta^{*}$ resonances need to be properly taken into account 
within the HRG inspired fits. If corrections due to pion-nucleon interactions 
are included within the HRG through the S-matrix formalism then the resultant 
statistical fit gives $T_f\sim 156$ MeV and at the same time correctly predicts
the proton yield \cite{Andronic:2020iyg}.

\begin{figure}[th]
\centerline{
\includegraphics[width=0.3\textwidth]{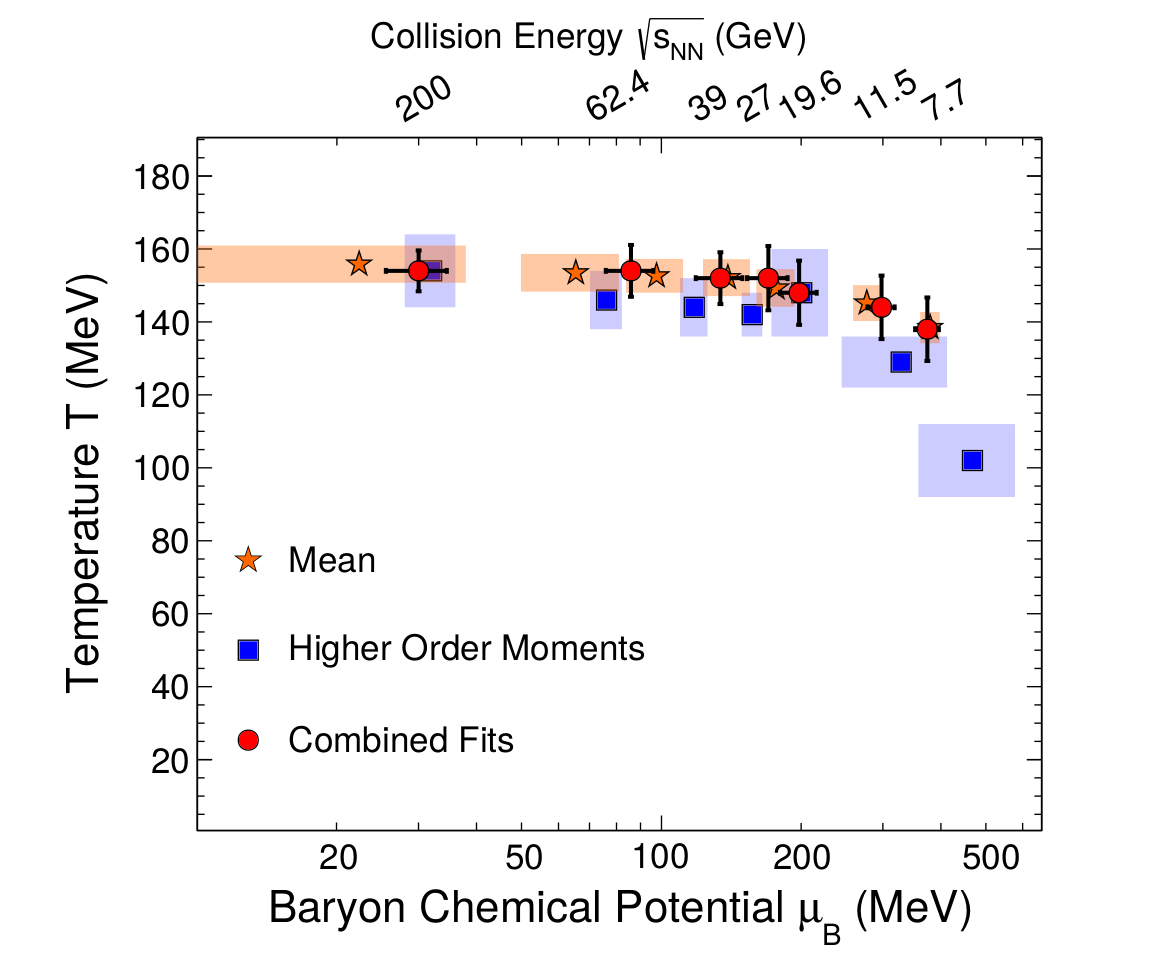}
\includegraphics[width=0.3\textwidth]{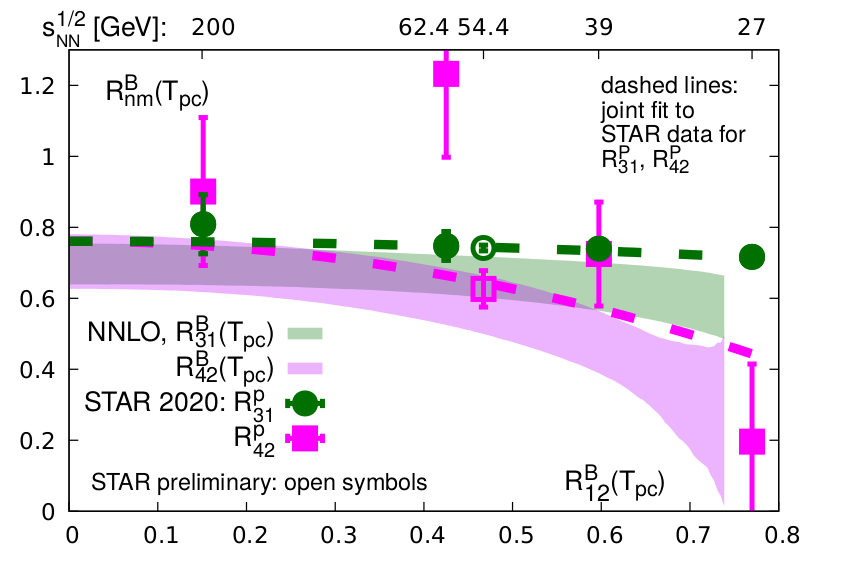}
\includegraphics[width=0.3\textwidth]{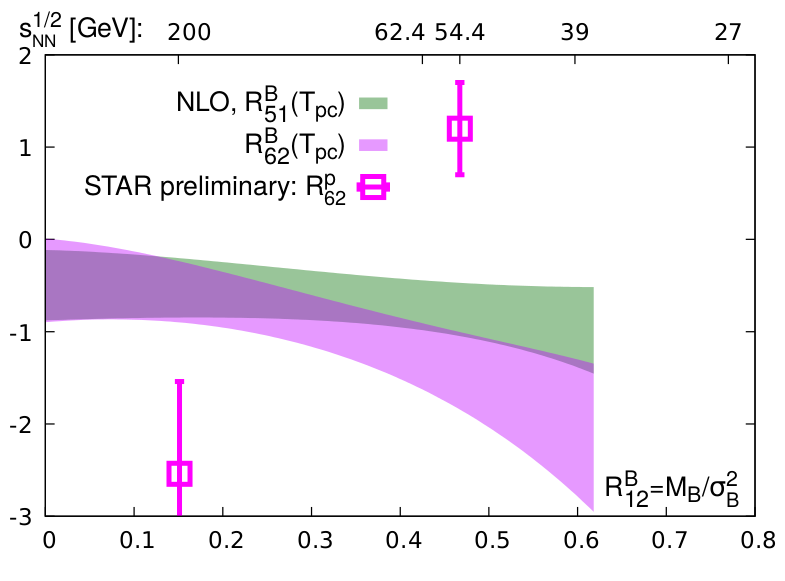}
}
\caption{The left panel shows the freezeout parameters as a function of 
$\sqrt s$ for RHIC energies obtained using improved statistical fits. 
The central and the right panels shows the comparison of lattice data on 
ratios of cumulants of baryon number fluctuations compared to proton number 
cumulants obtained at several $\sqrt s\geq 27$ GeV from STAR experiment. }
\label{fig:5}
\end{figure}

The phenomena of chemical freezeout has been studied extensively in the recent 
years. Few studies have discussed the possibility of multiple freezeout surfaces 
for the strange and non-strange hadrons \cite{Chatterjee:2013yga}, since the 
particle yields are not conserved quantities and hence yields of different 
species may not be described by same set of freezeout parameters, or in other 
words, may not attain simultaneous thermalization. 
A new proposal \cite{Gupta:2020pjd} for fitting the ratios of variances  
$\sigma^2_{QK}/\sigma^2_K$, $\sigma^2_{pK}/\sigma^2_K$, and $\sigma^2_{Qp}/\sigma^2_p$ 
along with $S\sigma_{p,K,Q}$ obtained from STAR (here $\sigma^2, S$ denotes the variance 
and skewness), gave consistent values of freezeout parameters as obtained only from the 
mean yields, for $\sqrt s \geq 19.6$ GeV. These results from Ref. \refcite{Gupta:2020pjd}
shown in the left panel of Figure \ref{fig:5}, demonstrate that thermalization could 
be achieved at the chemical freezeout. Constructing the fluctuations of various conserved 
charges from the experimental yields of hadrons, properly corrected for feed-down, 
and fitting them to corresponding quantities obtained within HRG also gave identical 
freezeout parameters with and without strangeness neutrality constraints 
\cite{Bhattacharyya:2019cer,Biswas:2020dsc}.

How far and if indeed thermal conditions are achieved at freezeout for the 
most central collisions can be checked by comparing suitable observables 
constructed from the experimental yields to those obtained from thermal 
QCD through lattice calculations. Systematic uncertainty remains in 
characterizing the freezeout parameters $T_f$ and the $\mu_B$ at each 
$\sqrt s$ and mapping it to the $T,\mu_B$ at which thermal QCD data 
are calculated \cite{Bazavov:2012vg}.  Moreover protons measured 
in the experiments are considered as a proxy for net-baryons
which introduces its own uncertainties in the analysis \cite{Kitazawa:2012at}. 
Away from the CEP, however such approximations may not be entirely wrong. 
It has been shown \cite{Bazavov:2012vg} that the ratio $\mu_B/T$ can be 
represented as the ratio of the mean ($M$) to the variance of net baryons 
at leading order since $R_{12}^B=M_B/\sigma^2_B=\mu_B/T+\mathcal{O}(\mu_B^3/T^3)$. 
This identification removes any systematic uncertainty in mapping the experimental 
freezeout parameters to thermal QCD data. Comparison of the ratio $S\sigma^3/M$ 
for the net proton yields from the most central collisions at STAR (here $S$ 
denotes the skewness of net proton), to $R_{31}^B=\chi_3^B/\chi_1^B$ calculated 
up to NNLO in $\mu_B/T$ on the lattice, as a function of $R_{12}^B$ from Ref.
 \refcite{Bazavov:2020bjn}, is shown in the central panel of Figure \ref{fig:5}. 
There is a striking consistency between these two observables. Similarly 
$R_{42}^p$ for a range of $\sqrt s$ except at $62.4$ GeV are consistent with 
the lattice data for $R_{42}^B$ \cite{Bazavov:2020bjn}. The latest preliminary 
STAR data for $R_{42}^p$ at $\sqrt s=62.4$ GeV is more closer to the lattice data. 
Ratios of yet higher order cumulants of baryon number measured on the lattice 
are noisy. It has been suggested \cite{Bazavov:2020bjn} to measure $R_{51}^p$ 
instead of $R_{62}^p$ and compare it to $R_{51}^B$ since the later can be 
measured with a good accuracy on the lattice. Consistencies between the 
lattice and experimental data on several such cumulant ratios cannot be 
a mere accident and will indeed lead to a better understanding of the 
phenomena of chemical freezeout.

\section{Outlook}
To summarize, lattice QCD at finite density is a fascinating area of research in 
theoretical sciences which will require some new ideas to solve the \emph{sign problem}.
 However even within such a constraint, new results have emerged circumventing this 
 problem, which gives very fundamental insights on non-perturbative QCD at finite 
 $\mu_B$ and also providing valuable inputs for the experimental efforts.   

\vspace{-0.5cm}
\section*{Acknowledgements}

I would like to acknowledge financial support from the Department of Science and 
Technology, Govt. of India through a Ramanujan Fellowship. I am also grateful to 
Frithjof Karsch for many helpful discussions.

\end{document}